\documentclass[prb,floatfix,showpacs,superscriptaddress,longbibliography,twocolumn]{revtex4-1}
\usepackage{amsmath,amssymb,amsfonts}

\usepackage{epsfig}
\usepackage{subfigure}
\usepackage[export]{adjustbox}

\usepackage[colorlinks=true,linkcolor=blue,citecolor=red]{hyperref}

\newcommand{\eq}[2]
{
  \begin{equation}
    #1
    \label{#2}
  \end{equation}
}

\newcommand{\eqsplit}[2]
{
  \begin{equation}
    \begin{split}
      #1
    \end{split}
    \label{#2}
  \end{equation}
}

\newcommand{\figu}[1]
{Fig.\ref{#1}}

\newcommand{\secu}[1]
{Sec.~\ref{#1}}

\newcount\colveccount
\newcommand*\colvec[1]{
  \global\colveccount#1
  \begin{pmatrix}
    \colvecnext
  }
  \def\colvecnext#1{
    #1
    \global\advance\colveccount-1
    \ifnum\colveccount>0
    \\
    \expandafter\colvecnext
    \else
  \end{pmatrix}
  \fi
}

\newtoks\rowvectoks
\newcommand{\rowvec}[2]{%
  \rowvectoks={#2}\count255=#1\relax
  \advance\count255 by -1
  \rowvecnexta}
\newcommand{\rowvecnexta}{%
  \ifnum\count255>0
  \expandafter\rowvecnextb
  \else
  \begin{pmatrix}\the\rowvectoks\end{pmatrix}
  \fi}
\newcommand\rowvecnextb[1]{%
  \rowvectoks=\expandafter{\the\rowvectoks&#1}%
  \advance\count255 by -1
  \rowvecnexta
}

\def\bcen{\begin{center}}
\def\ecen{\end{center}}

\def\a{\alpha}

\def\p{\pi}

\def\CC{{\cal C}}

\def\=={\equiv}

\def\qed{\raise1pt\hbox{\vrule height5pt width5pt depth0pt}}

\def\cG0{{\cal G}_0}
\def\cG{{\cal G}}

\def\up{\uparrow}
\def\down{\downarrow}
\def\dw{\downarrow}

\def\ka{{\bf k}}

 \def\Im{\mbox{Im}}

\def\=={\equiv}

\def\Im{{\rm Im}}
\def\Re{{\rm Re}}
\def\Tr{{\rm Tr}\,}

\def\ep0{\epsilon_{p}}
\def\ed0{\epsilon_{d}}

\def\ka{{\bf k}}
\def\kx{{ k_x}}

\def\kxi{{k_x\!y}}

\def\yi{{ y}}
\def\ia{{\bf i}}
\def\ja{{\bf j}}

\def\opN{\hat{ N}}
\def\opSz{\hat{ S}_z}
\def\opTz{\hat{ T}_z}
\usepackage{bbold}
\def\11{\mathbb{1}}
\def\00{\mathbf{0}}

\begin{document}
\title{Coexistence of metallic edge states and anti-ferromagnetic ordering in correlated topological insulators}

\author{A.~Amaricci}
\affiliation{Scuola Internazionale Superiore di Studi Avanzati (SISSA),
and  Consiglio Nazionale delle Ricerche,
Istituto Officina dei Materiali (IOM),  Via Bonomea 265, 34136 Trieste, Italy}

\author{A.~Valli}
\affiliation{Scuola Internazionale Superiore di Studi Avanzati (SISSA),
and  Consiglio Nazionale delle Ricerche,
Istituto Officina dei Materiali (IOM),  Via Bonomea 265, 34136 Trieste, Italy}

\author{G.~Sangiovanni}
\affiliation{Institut f\"ur Theoretische Physik und
  Astrophysik, Universit\"at W\"urzburg, Am Hubland, D-97074 W\"urzburg, Germany}

% \author{J.~C.~Budich}
% \affiliation{Max-Planck-Institut f\"ur Physik komplexer
%   Systeme, N\"othnitzer Stra{\ss}e 38, 01187-Dresden, Germany}

\author{B.~Trauzettel}
\affiliation{Institut f\"ur Theoretische Physik und
  Astrophysik, Universit\"at W\"urzburg, Am Hubland, D-97074 W\"urzburg, Germany}

\author{M.~Capone}
\affiliation{Scuola Internazionale Superiore di Studi Avanzati (SISSA),
and  Consiglio Nazionale delle Ricerche,
Istituto Officina dei Materiali (IOM),  Via Bonomea 265, 34136 Trieste, Italy}

\date{\today}

\begin{abstract}
We investigate the emergence of anti-ferromagnetic ordering and its
effect on the helical edge states in a quantum spin Hall insulator, in the
presence of strong Coulomb interaction. Using dynamical mean-field
theory, we show that
%that in presence of electronic interaction
the breakdown of lattice translational symmetry favours the formation of
magnetic ordering with non-trivial spatial modulation.
The onset of a non-uniform magnetization enables the coexistence of
%the ordered state with a persisting topological state.
spin-ordered and topologically non-trivial states.
An unambiguous signature of the persistence of the topological bulk
property is the survival of bona fide edge states.
We show that the penetration of the magnetic order is
accompanied by the progressive reconstruction of gapless states in sub-peripherals
layers, redefining the actual topological boundary within the system.
\end{abstract}


\maketitle

\section{Introduction}
Understanding the impact of  strong electron-electron interactions on Topological
Insulators (TIs) is an open
challenge~\cite{Lehur10,Hohenadler2011,Hohenadler2012,Tada2012PRB,Budich2012PRB,Budich2013PRB,Hohenadler2013JOPCM,xHung2014PRB,xLu2013PRL,Amaricci2015PRL,Amaricci2016PRB,Roy2016PRB,Kumar2016PRB}
which holds the potential for the discovery of novel states of matter.
One of the most subtle issues is that, while intrinsic
electron-electron correlation effects may have a non-trivial interplay
with topological properties, strong interactions typically  favour
different kinds of spontaneous symmetry breaking and the formation of
long-range-ordered states (e.g. spin, charge or orbital ordering).

The paradigmatic case for non-frustrated correlated insulators is the anti-ferromagnetic (AFM)
ground-state resulting from a local Hubbard repulsion.
If we consider TIs, we can expect the magnetic ordering to obliterate the non-trivial
topological character because of the breaking of the
time-reversal symmetry (TRS).~\cite{Kane2005PRL,Kane2005PRLa,Bernevig2006S}
Suitable topological properties can however be preserved if an alternative or
a residual symmetry persists.
For commensurate AFM states a symmetry playing the
role of time-reversal can be realized by combining complex conjugation
and lattice translations.~\cite{mongPRB81,fangPRB88}
Such symmetry gives rise to a ${\bf Z}_2$-classification in
three dimensions but not in two.

The interplay between topological properties and correlation-driven AFM
acquires further depth in a finite system, whose boundaries can host peculiar edge states, and
at the same time they show stronger electron correlation effects~\cite{Shitade2009PRL,Borghi2009PRL,Borghi2010PRB,Mazza2015PRB,Medhi2012PRB,Ishida2014PRB,AmaricciPRB95}.
In this article, we explore this situation by studying  AFM ordering in the Bernevig-Hughes-Zhang (BHZ)
model,~\cite{Bernevig2006S} supplemented by a local
interaction~\cite{Budich2013PRB,Amaricci2015PRL,AmaricciPRB95} on a
stripe  with open boundaries along one spartial direction using Dynamical Mean-Field Theory
(DMFT).~\cite{Georges1996RMP,Kotliar2006RMP}

For a strong enough interaction the system undergoes
a first-order transition to an AFM state.
Near the transition the magnetization along the confined direction deviates from a simple
staggered pattern, allowing for a coexistence with a persisting topological state.
Surprisingly, the helical edge states~\cite{Konig2007S,Roth09,Knez11} survive in the presence of AFM
ordering for a large window of the interaction strength.
When the external layers become insulating the edge states penetrate the ``sub-peripheral'' layers,
redefining the actual boundaries of the TI.
This edge {\it reconstruction} is similar to that taking place near a
Mott insulator~\cite{Medhi2012PRB,Ishida2014PRB,AmaricciPRB95} which,
however, does not break any of the symmetries protecting the
topological state.

The novel boundary layers not only separates the topological bulk
from the trivial ordered region,
but also marks a dramatic change in the many-body character of the system.
While the insulating AFM layers can be basically described in terms of
Hartree-Fock theory,~\cite{sangiovanniPRB73} the inescapable presence
of the gapless edge states turns out to bring correlation effects,
beyond Hartree-Fock picture, back into the game. 
The outcome of correlations, topology and boundary effects is therefore a strongly correlated
AFM state which remains metallic with gapless edge states.

The topological nature of the system arises from the residual $U(1)$ symmetry,
corresponding to the spin-rotation around the $z$-axis,
as the magnetic ordering does not have in-plane components.
In general, the $U(1)$-invariant topological states are referred
to as spin-Chern insulators (SCIs).~\cite{yoshidaPRB87,chenPRB91,rachelReview}
The emergence of an interaction driven SCI in the bulk have been
addressed before using different approaches~\cite{Lehur10,yoshidaPRB87,miyakoshiPRB87}.
However, the analysis of the corresponding edge states in a finite system is still lacking.
Our calculations demonstrate that a SCI is realized in a finite
geometry near a correlation driven AFM transition.

The article is organized as follows. In
\secu{Sec1}, we introduce the interacting BHZ model (that allows for magnetic ordering) and discuss the method of
solution. In \secu{Sec2}, we present the main results of our
work. In particular, we investigate the antiferromagnetic ordering in a stripe geometry with distinction of bulk and edge properties. Finally, in \secu{Sec3}, we conclude and point out interesting perspectives.

\begin{figure}
  \includegraphics[width=0.49\textwidth]{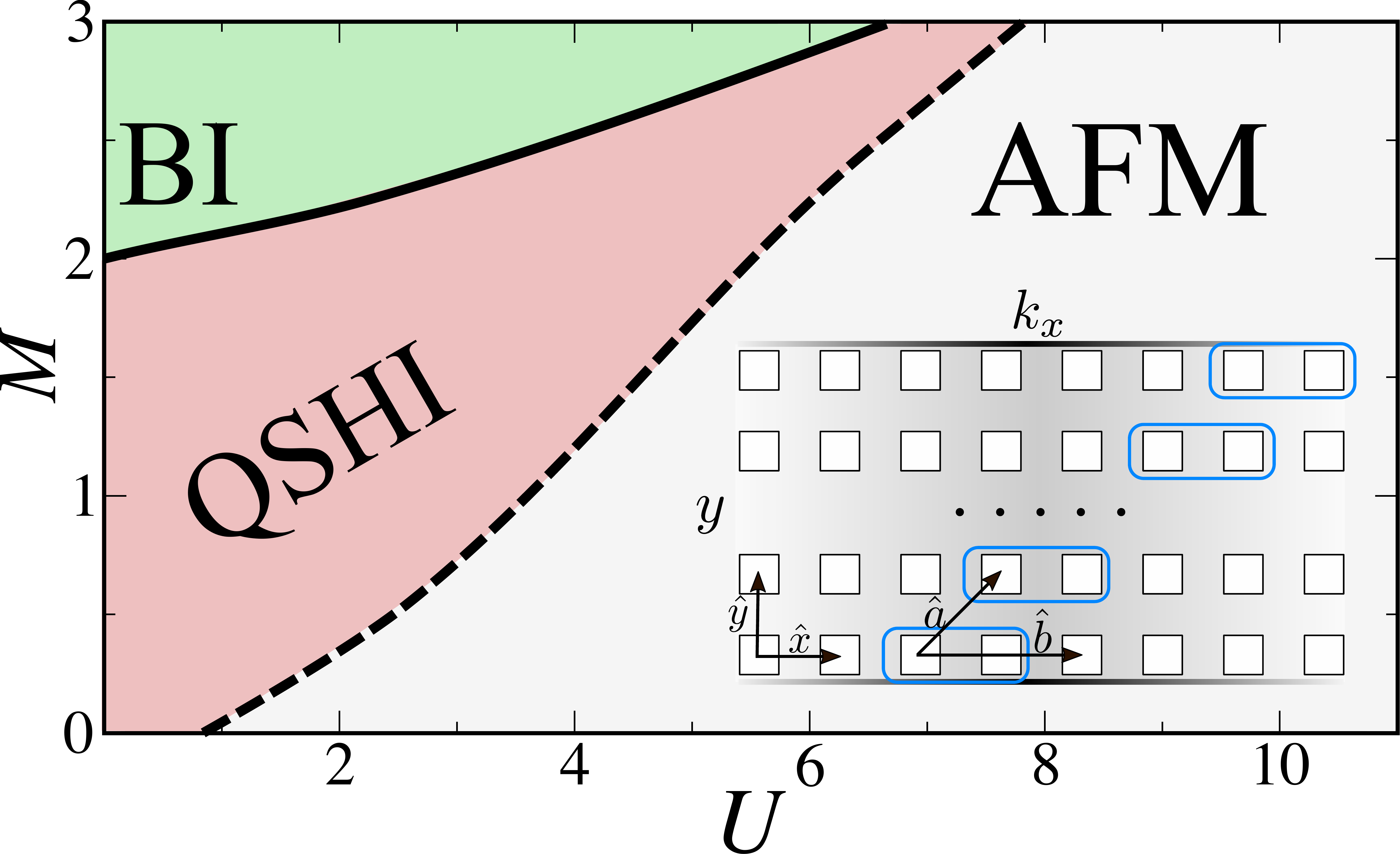}
  \caption{
    (Color online) {\it Main panel}: $M$-$U$ phase diagram of the model 
    with periodic boundary conditions in both directions.
    For small $U$, the system is either a trivial band insulator (BI)
    or a quantum spin Hall insulator (QSHI).
    For larger interaction strength, an AFM solution becomes stable.
    The ordered phase is separated from
    the QSHI by a first-order transition.
    {\it Inset}: schematic structure of the main model in a stripe
    geometry with a unit cell hosting two inequivalent sites.
  }
  \label{fig1}
\end{figure}
\section{Interacting BHZ Model}\label{Sec1}

We consider a two-orbital BHZ
model in two dimensions in presence of a local
interaction term.~\cite{Budich2013PRB,Amaricci2015PRL,AmaricciPRB95}
The Hamiltonian can be written in terms of the identity $\Gamma_0
\!=\! \11 \otimes\11 $ and the following 4$\times$4 matrices:
$\Gamma_x \!=\! \sigma_z \otimes \tau_x$,
$\Gamma_y \!=\!-\11 \otimes \tau_y $,
$\Gamma_5 \!=\! \11 \otimes \tau_z$ and
$\Gamma_\sigma\!=\!\sigma_z\otimes\11$.
Here, $\sigma_{x,y,z}$ and $\tau_{x,y,z}$ are two sets of Pauli
matrices acting, respectively, on the spin and orbital sector.

If periodic boundary conditions (PBC) are assumed in both spatial 
directions, the non-interacting part of the Hamiltonian can be diagonalized in momentum space:
$H = \sum_\ka \psi_\ka^\dag { H}_\ka\psi_\ka$,                      
where the spinor $\psi_{\ka=(k_x,k_y)}$
collects the operators
$c_{\ka \a\sigma}$ destroying electrons at the orbital $\a=1,2$ with
spin $\sigma=\up,\dw$, while
${ H}_\ka=
 E(\ka)\Gamma_5 +  \lambda \sin(k_x)\Gamma_x +
 \lambda \sin(k_y)\Gamma_y$ and
$E(\ka)\!=\!M\!-\!\epsilon[\cos(k_x)\!+\!\cos(k_y)]$.
This Hamiltonian is invariant under TRS and conserves $S_z$,
i.e. it has $U(1)$ spin rotation symmetry.

\begin{figure}
  \includegraphics[width=0.49\textwidth]{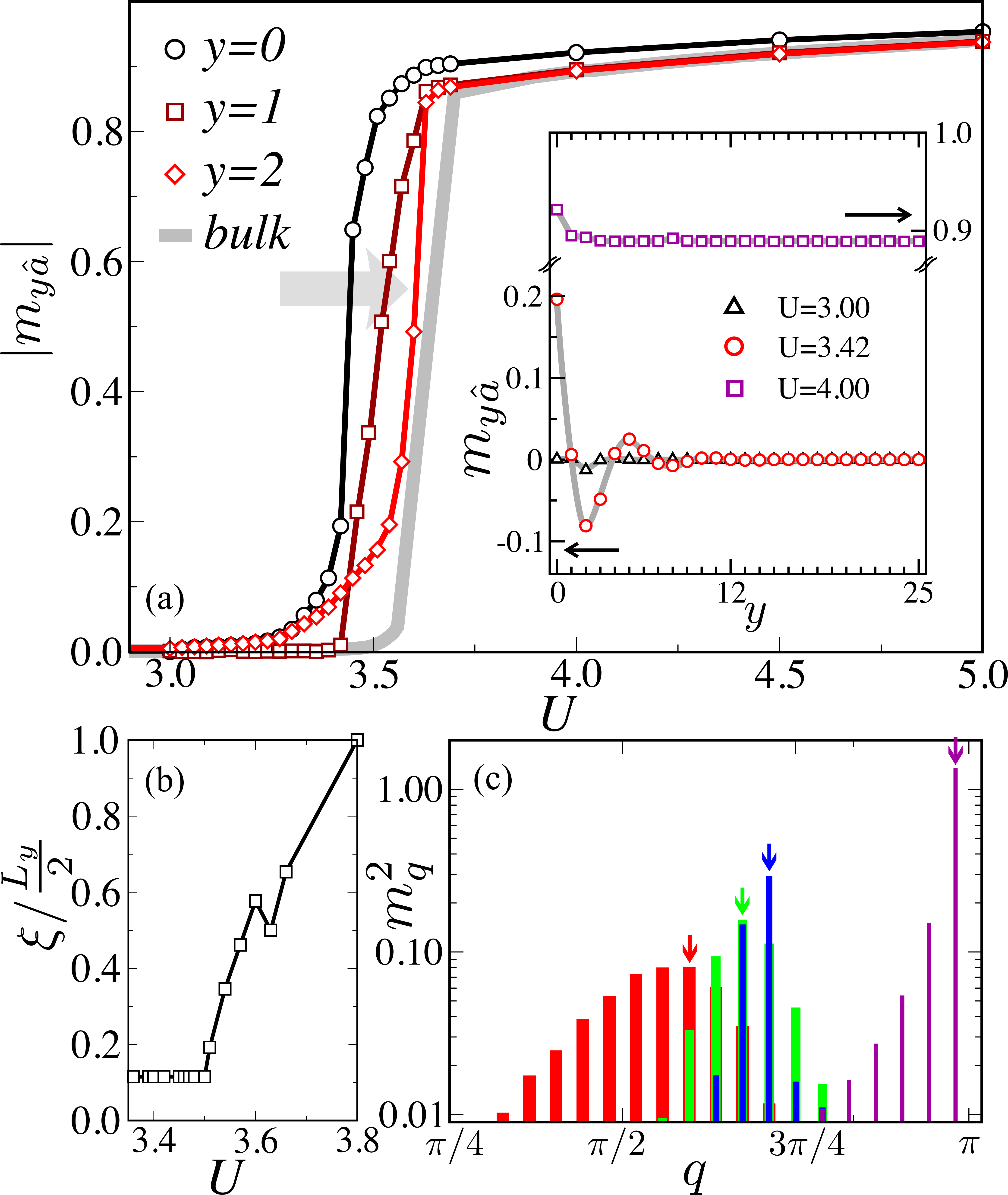}
  \caption{(Color online) {\it (a)}: Magnetization $|m_{y\hat{a}}|$
    along the $\hat{a}$ direction as a function of $U$ and for
    the first three layers $y=0,1,2$. Data are shown for the choice of site $\Lambda=A$.
    The thick (gray) line corresponds to the magnetization behavior in
    the homogeneous case.
    {\it Inset}: Layer resolved magnetization $m_y$ for $\Lambda=A$
    and three values of the interaction strength $U$.
    {\it (b)}: Correlation length of the magnetization profile
    $\xi$ as a function of $U$. Data are in units of the half size of
    the stripe $L_y$ along the transverse direction.
    {\it (c)}: Evolution of the Fourier spectrum of the
    magnetization $m^2_{q}$ across the AFM transition.
  }
  \label{fig2}
\end{figure}

In order to describe the effects of AFM ordering, we consider
a unit cell with two inequivalent sites, $\Lambda=A,B$ (see inset in \figu{fig1}).
The unit cell tiles the lattice along the two vectors
$\hat{a}=a_{lat}(1,1)$, $\hat{b}=a_{lat}(2,0)$, with $a_{lat}$ is the lattice constant which we assume as our unit length.

In this paper we solve the model in a geometry where we have open boundaries and a finite number $L_y$ of layers along the $\hat{a}$ direction, while PBC are assumed along the $\hat{b}$ direction  ("stripe geometry") as shown in the inset of \figu{fig1}.
In the hybrid basis $(k_x,y)$, the model Hamiltonian takes the form
\eqsplit{
  H = &  \sum_{\kx,\yi,\yi'}
  \Psi^\dag_\kxi \mathbf{M}(\kx) \delta_{y y'} \Psi_{\kx\!y'} + \cr
  & \sum_{\kx\yi,\yi'}  \left(
  \Psi^\dag_\kxi \mathbf{A}(\kx) \delta_{y\!+\!1 y'} \Psi_{\kx\!y'} \! +\!
  H.c.\right)
  + H_\mathrm{int} \; ,
}{model_hamiltonian}
where $\yi\!=\!0,\dots,L_y-1$ is the coordinate in $\hat{y}$
direction, i.e. the layer index, $\Psi_\kxi=(\psi_{\ka,A},\psi_{\ka,B})$, and
\begin{equation*}
  \begin{split}
    \mathbf{M}(\kx) & \!=\!
    \begin{bmatrix}
      \hat{m} & \hat{t}_x + \hat{t}_x^\dag e^{i 2k_x}\\
      \hat{t}_x^\dag + \hat{t}_x e^{-i 2k_x} & \hat{m}
    \end{bmatrix} \; , \\
    \mathbf{A}(\kx) & \!=\!
    \begin{bmatrix}
      \hat{0} & \hat{t}_y^\dag e^{ik_x}\\
      \hat{t}_y^\dag e^{-ik_x} & \hat{0}\\
    \end{bmatrix}\\
  \end{split}
\end{equation*}
with $\hat{m}=M\Gamma_5$,
$\hat{t}_x=-\frac{\epsilon}{2}\Gamma_5 + i\frac{\lambda}{2}\Gamma_x$,
$\hat{t}_y=-\frac{\epsilon}{2}\Gamma_5 + i\frac{\lambda}{2}\Gamma_y$.
In the remainder of this article, we set $\epsilon$ as the energy unit.

In Eq.~(\ref{model_hamiltonian}), $H_\mathrm{int}$ describes a local Coulomb interaction with both
{\it inter}- and {\it intra}-orbital repulsion and the Hund's coupling
$J$, taking into account the exchange effect which favors  high-spin configurations.
In terms of local operators,
$\opN\! =\!\sum_{\ia\ja}\Psi^\dag_{\ia}\Gamma_0\delta_{\ia\ja}\Psi_{\ja}$,
$\opSz\!=\!\tfrac{1}{2}\sum_{\ia}\Psi^\dag_{\ia}\Gamma_\sigma\delta_{\ia\ja}\Psi_{\ja}$,
$\opTz\!=\!\tfrac{1}{2}\sum_{\ia}\Psi^\dag_{\ia}\Gamma_5\delta_{\ia\ja}\Psi_{\ja}$,
the interaction term reads
\eq{
  H_\mathrm{int} = (U-J)\frac{\opN(\opN-1)}{2}
- J \left(  \frac{\opN^2}{4} + \opSz^2 - 2\opTz^2 \right)\, ,
}{model_interaction}
where $U$ is the strength of the electron-electron interaction and
 $\Psi_{\ia=x,\yi}=\sqrt{\tfrac{2\p}{V}}\sum_\kx e^{-i\kx\cdot
   x}\Psi_\kxi$
\footnote{This Hamiltonian only contains the ``density-density'' part of the Hund's
exchange and neglects the so-called pair-hopping and spin-flip
terms. The robustness of the topological transitions in the BHZ
model against the pair-hopping and spin-flip terms has been
verified in Ref.~\onlinecite{Budich2013PRB}}.
In the following,
we fix $\lambda = 0.3$ and a relatively large ratio $J\!=\!U/4$ but none of our results are specific to this choice.

We solve the interacting problem non-perturbatively using
DMFT~\cite{Capone2007PRB,Weber2012PRB,Amaricci2015PRL,Amaricci2016PRB,AmaricciPRB95} at zero temperature.
In order to capture the different behavior between bulk and
boundaries, we rely on its real-space extensions for inhomogeneous
systems.~\cite{yPotthoff1999PRB,yFreericks2006,Snoek2008NJP,yAmaricci2014PRA,Valli2012PRB,Valli2016PRB,Valli2018NL}
In this framework, the interaction effect is contained in
a diagonal but layer-dependent self-energy function
$\mathbf{\Sigma}_y(\omega)$, bearing the correct spin-orbital structure.

\section{AFM ordering and metallic edge states}\label{Sec2}
The presence of strong electronic interaction usually favours the
emergence of an instability towards long-range ordered states at low temperature. 
For half-filled bipartite lattices the leading instability is towards a staggered AFM ordering.
Our system makes no exception to this generic consideration.
We clarify this point in \figu{fig1},  reporting the phase diagram of the
model in an infinite system with PBC in both directions.~\cite{Amaricci2015PRL}
In the non-interacting regime ($U=0$) the system undergoes a
topological quantum phase transition at $M=2$ from a band-insulator (BI) for $M>2$
to a quantum spin Hall insulator (QSHI) at $M<2$.
These two phases are adiabatically continued into separate regions of
the diagram for finite values of the interaction.
The QSHI region is topologically non-trivial hence the ${\bf Z}_2$
invariant is   $\nu=(\CC_\up - \CC_\dw)/2 = 1$, where $\CC_\sigma$ is
the Chern number for a given spin orientation.~\cite{Hohenadler2013JOPCM}
In both the QSHI and the BI regions the ground state is paramagnetic and TRS is preserved.
However, for larger values of the interaction strength the system
undergoes a transition from the paramagnetic QSHI to an AFM phase.
The magnetic transition takes place at  $U=U_c^{homo}$
and it is of the first order (see \figu{fig2}(a)).
In the homogeneous case the AFM phase is topologically trivial as TRS is explicitly broken.
Notably, we cannot exclude the possible existence of a SCI phase
in the coexisting region near the critical point, as pointed out in Ref.~\onlinecite{Yoshida2013PRB}.
The very existence of such non-trivial metastable phase requires a
fine-tuning of the model parameters. 
We will show below that an 
extended SCI phase, characterizing by a finite decaying magnetization
in the bulk, naturally emerges in a finite geometry as a result of
strong correlations.

% \gs{Here we have to describe the AFM phase. IMHO it would be nice to
%   understand whether the properties of the distribution in $q$ are a
%   consequence of the existence of the topological edge states. This
%   may be a strong selling point for the paper. I would also show a
%   figure with a close-up of the edge states where we can show that
%   $\epsilon_\uparrow(k) \neq \epsilon_\downarrow(-k)$ due to TRS
%   breaking, for each edge separately.}

\begin{figure*}
 \includegraphics[width=1\textwidth,left]{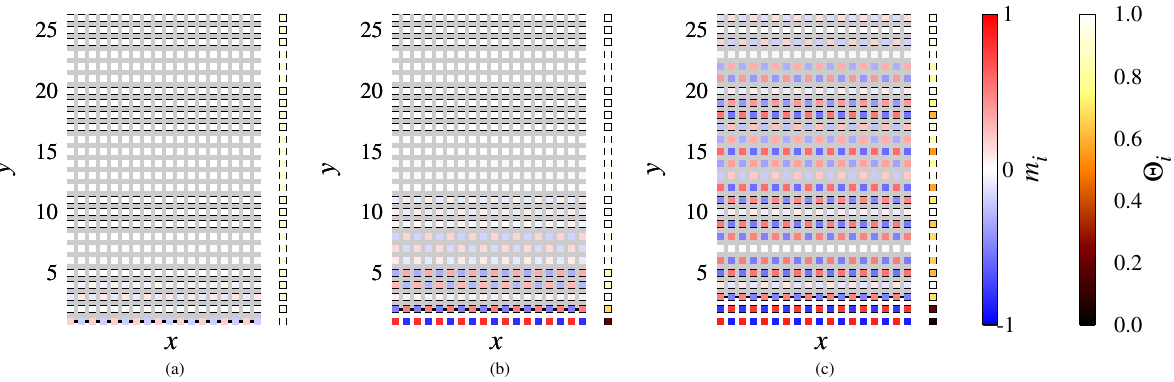}
  \caption{
    (Color online) Spatially resolved local magnetization $m_i$
    and correlation strength $\Theta_i$ in the stripe
    for increasing values of the interaction strength across the AFM transition,
    $U=3.42$ (a),  $3.51$ (b) and $3.60$ (c). The figures
    correspond to half of the stripe. Data are displayed for $L_y=52$.
    The color scale for the magnetization and the correlation strength
    are reported on the vertical bars on the right. The (black)
    solid lines indicate the location of the reconstructed edge states.
  }
  \label{fig3}
\end{figure*}
We now consider the system on the stripe geometry introduced
above. To characterize the magnetic nature of the solution we consider the local magnetization
$m_{\ia}=\tfrac{1}{2}\sum_\alpha (n_{\ia\a\up}-n_{\ia\a\down})$, where $n_{\ia\a\sigma}$ is the
local occupation at the site $\ia$, for the orbital $\a$ with
spin $\sigma$.
In a finite system, the lattice translational symmetry is broken, leading to
a substantial reduction of the kinetic energy near the
boundaries. This makes the sites near the edge effectively more
correlated.~\cite{Borghi2009PRL,Borghi2010PRB,Valli2012PRB,AmaricciPRB95}
Hence, we expect the external layers to build up a magnetic
order more rapidly with respect to the bulk as the interaction
increases.~\cite{Valli2016PRB}

This effect is illustrated in \figu{fig2}(a) here we plot  the
layer-resolved absolute value of the magnetization $|m_{y\hat{a}}|$,  along the
diagonal direction $\hat{a}$ for the sublattice $\Lambda=A$. This quantity would 
be a constant as a function of $y$ in an ideal N\'eel antiferromagnet.
Here we plot the absolute value to focus on the way the interaction leads to the formation
of magnetic moments in the first layers.
As a function of $U$, the initially non-magnetic system at weak coupling is progressively
driven into an ordered state by increasing the interaction
strength. However, the most external layer, $y=0$,
reaches the saturated value for the magnetization for smaller interactions than the more internal
layers. The behavior of $m_{y\hat{a}}$ shows a sequential behavior
with respect to the different layers, near the saturation point.
In this regime, few external magnetic layers
coexist with a paramagnetic ``bulk".
Further increasing the interaction strength, the AFM ordering
penetrates the inner layers. For interactions larger than a critical
value $U>U^{stripe}_c\simeq U^{homo}_c$ the whole system becomes AFM.

Interestingly, the spatial distribution of the
magnetization has a non-trivial behavior which we explore
In the inset of \figu{fig2}(a).
Here we report the evolution of the magnetization
$m_{y\hat{a}}$ as a function of the layer index $y$ along the
$\hat{a}$-direction and for the sublattice $\Lambda=A$. Along the periodic direction
a perfect N\'eel ordering is imposed.

For weak interactions, the magnetization is negligible
everywhere reflecting the non-magnetic nature of the solution in this regime.
By increasing the correlation strength  in the regime $U<U_C^{stripe}$,
%the system develops a finite magnetization near the boundaries which,
%in a finite system away from the transition point,
the system develops a finite magnetization near the boundaries which
is exponentially suppressed in the bulk on a length scale $\xi$.
Interestingly, the magnetization shows a non-trivial spatial modulation before the staggered pattern
(which corresponds to a constant magnetization with our notation) sets in.
In this regime, the ordered boundary layers coexist with a yet non-magnetic bulk.
Only for a larger interaction strength, the system eventually develops a
nearly constant and flat magnetization profile $m_{y\hat{a}}$,
corresponding to the complete transition of the stripe to an AFM state.

In order to get insight into the development of the AFM state and the
non-trivial spatial distribution of the order parameter, we
investigate both the evolution of the correlation length in transverse
direction
$\xi$ and the Fourier decomposition of the magnetization profile,
i.e. $m_{q}=\sqrt{\frac{2\pi}{L_y}}\sum_{y} e^{iqy}m_y$
for $q=\frac{n\pi}{L_y}$ and $n=0,\dots,L_y-1$.
The results for these two quantities are reported in \figu{fig2}(b)-(c).
The following scenario emerges.
Approaching the AFM transition, the correlation length increases from
few ($\xi\simeq2$) to many lattice units.
Thus, enhancing the interaction strength,
an increasing number of layers
%more and more layers
from the boundary to the bulk order magnetically.
When the correlation length reaches the half-size of the stripe, $\xi\simeq L_y/2$,
the two AFM fronts ``propagating" from the opposite boundaries overlap.
Beyond this point the system becomes entirely AFM.
Simultaneously, the spectral distribution of the
order parameter undergoes a dramatic change.
Initially, the distribution has a wide ``bell''-like shape, spread over many momenta and
%it is
centered near $q_{peak}\simeq\tfrac{\pi}{2}$.
Increasing $U$ the width of the distribution reduces
while the peak contribution increases its weight.
Approaching the transition the spectral
distribution of the AFM order parameter becomes entirely dominated by the peak contribution,
which in turn approaches the value $q_{peak}=\pi$.
Eventually, at $U>U_c^{stripe}$
the magnetization converges towards the form $m_\ia = e^{i{\bf Q}\ia}
M$, where ${\bf i}=(x,y)$ and ${\bf Q}=(\pi,\pi)$ is the ordering
vector of the N\'eel AFM state, as expected in the homogeneous limit,
i.e. PBC in all directions.

The construction of the AFM state in the stripe comes with
the progressive breakdown of TRS. In the absence of such symmetry
the conventional QSHI is modified and its helical states at the boundary
change nature.
However, we found that near the AFM transition the
system remains in a topological state, endowed with vestigial
edge states. Such gapless modes are protected
by the residual $U(1)$ symmetry associated with the spin-rotation around the easy
axis. If this symmetry was broken, a gap would open
for the edge states.
The presence of the $U(1)$ symmetry is enough to
stabilize the topological state against the magnetic ordering,
leading to a progressive {\it reconstruction} of the edge
states,~\cite{AmaricciPRB95} and the consequent {\it contraction} of the
topological bulk, as the AFM order penetrates into the system.
In order to illustrate this effect we report in \figu{fig3}(a)-(c) the evolution
of the spatial distribution of the magnetization and of the
reconstructed edge states.
For small values of the interaction strength the staggered
magnetization is weak and localized at the boundary layer, where it
coexists with the gapless mode.
As the interaction is increased the AFM order slowly builds up and
progressively penetrates into the stripe. As the magnetization reaches
its maximum value at the most external layer, a gap in the energy
spectrum opens and the gapless edge states shifts inward.
In this regime of the interaction strength $U<U_c^{stripe}$, the system hosts a
non-trivial topological bulk with a residual weak magnetization
distribution, separated from the trivial AFM part at the
most external layers by a conductive state located in
``sub-peripherals'' layers.

The discontinuity in the topological character at
the edge layer is accompanied by a dramatic change in the nature
of the ground state.
The topologically non-trivial bulk shows a strongly correlated nature.
We estimate the correlation strength in terms of the electronic
self-energy as:
$$
\Theta_{\bf i} = \frac{\Tr\left[\Gamma_5\Re{\bf \Sigma}_{\bf i}(0)  -
\Gamma_5{\bf \Sigma}^{\rm \scriptsize HF}_{\bf i}\right]}
{\Tr\left[\Gamma_5\Re{\bf \Sigma}_{\bf i}(0)\right]}
$$
where ${\bf \Sigma}^{\rm \scriptsize HF}_{\bf
  i}=\lim_{n\rightarrow\infty}\Re{\bf \Sigma}_{\bf i}(i\omega_n)$.
%On the other hand, within the DMFT solution the AFM state is well
%described in terms of a Hartree-Fock theory. The resulting
%ground state has a low degree of electronic correlation, i.e. has
%a single-particle nature.
%Remarkably, near the AFM transition the penetration of the magnetic
%distribution inside the topological bulk carries correlations effects
%beyond the expected Hartree-Fock portrait of the ordered state.
Deep inside the AFM ordered state, the DMFT yields very similar results
to the static Hartree-Fock theory, and the ground state
has a low degree of electronic correlation, i.e. the electronic states are very close to single-particle states~\cite{sangiovanniPRB73}
Near the transition,
the penetration of the AFM ordering in the strongly interacting topological bulk
gives rise to a state that is simultaneously correlated and
magnetic. As such, the nature of this state goes beyond a conventional
Hartree-Fock description.
In order to illustrate this point we show in \figu{fig3} the behavior of $\Theta_{\bf i}$
on one half of the stripe along the transverse direction.
%For smaller interactions the correlation strength is nearly at its maximum everywhere
%Approaching the magnetic transition, $\Theta_{\bf i}$ gets suppressed at the ordered
%layers. The correlation strenght however, remains large in the
%topological bulk, despite the presence of a
%finite magnetization which in turn determines the spatial distribution
%of $\Theta_{\bf i}$.
At $U<U_c^{stripe}$ the correlation strength is nearly at its maximum at all layers.
Approaching the magnetic transition,
$\Theta_{\bf i}$ is strongly suppressed in the trivial AFM boundaries,
and it remains finite at the reconstructed edges
as well as in the topological bulk, despite the progressive penetration of magnetization.
Remarkably, we observe the same spatial modulation
between the magnetization and the suppression of $\Theta_{\bf i}$.
The system retains a substantial degree of correlation
until the AFM correlation length reaches the value $\xi\sim L_y/2$,
and the whole system becomes  a trivial AFM insulator.

\begin{figure}
 \includegraphics[width=0.475\textwidth]{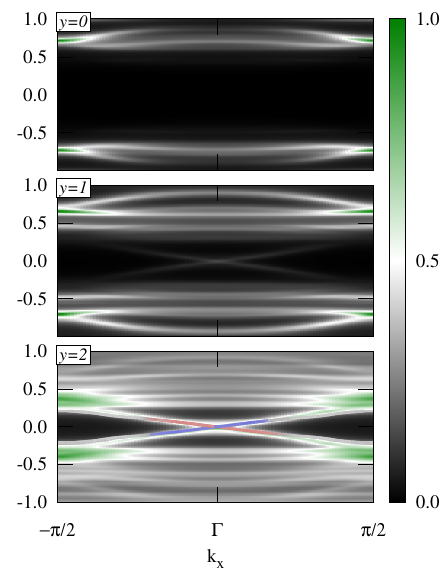}
  \caption{
    (Color online) Layer- and $k_x$-resolved spectral
    functions $A_y(k_x,\omega)$ for $y=0$ (top), $y=1$ (center) and $y=2$
    (bottom). Data are for $U=3.60$. The solid lines indicate the
    dispersion of the gapless edge states located at the third layer.
  }
  \label{fig4}
\end{figure}

Finally, we can characterize the edge reconstruction phenomena
studying the spatial evolution of
the $k_x$- and layer-resolved spectral function:
$A_y(k_x,\omega) =
-\tfrac{1}{4\pi}\sum_{\alpha,\sigma}\Im{G}_{\alpha,y}(k_x,\omega+i0^+)$.
In \figu{fig4}, we report the low-energy part of $A_y(k_x,\omega)$
around the $\Gamma$ point
for the first three layers at one edge of the stripe, near the AFM transition.
For $U=3.60<U_c^{stripe}$, the gapless helical states
at the $y=0$ layer have collapsed leaving behind a spectral gap
associated to the AFM ordering.
While some sub-gap spectral weight is developed, due to hybridization
with the bulk layers, in the second interior layer ($y=1$), it is only at the third layer ($y=2)$
that we can find a new pair of gapless states,
renormalized by electronic correlations separating the
trivial AFM insulator from the topologically non-trivial bulk.
The situation is identical at the opposite side of the stripe.

% The topological nature of the bulk near the AFM transition can be inferred by
% looking at the properties of the reconstructed edge
% states. This is done in \figu{fig3}(g) where we report the
% renormalized low-energy band structure of the system. A
% characterization of the
% spin, orbital and layer properties of the gapless edge states point out
% the SCI nature of the bulk.

\section{Conclusions and Outlook}\label{Sec3}
We have investigated the effect of AFM ordering caused by strong local interactions
on the topological insulating state of the BHZ model and in particular on the 
gapless edge states.

Solving the model on a stripe with open boundaries, we have pointed out the existence of a first-order
transition to a N\'{e}el type AFM state.
The transition is preceded by a regime in which the magnetic order
has a non-trivial spatial modulation.
Unlike in conventional insulators, the onset of inhomogeneous magnetic order does
not destroy entirely the topological insulating state which, instead, survives in the bulk.
The opening of a magnetic gap in the most external layer is
accompanied by the creation of novel helical edge states in the immediately adjacent layers,
which become the actual topological boundary of the system. These edge states are
protected by a residual $U(1)$ symmetry.
Thus, our findings unveil a rich phenomenology of the edge states of
interacting spin-Chern insulators.
In particular our calculations suggest a novel way to induce a metallic state
in an antiferromagnet caused by strong electronic correlations.

A conventional way to drive an AFM into a
metallic state is through hole doping. This comes with a number of fascinating consequences, such
as spin-charge separation in one dimension and confined spin polarons. It is even believed to cause high-temperature superconductivity in two dimensions. Yet, injecting holes in an AFM state inevitably spoils the
long-range order and eventually destroys it.
In our work, we find that symmetry protection of the edges in a topologically
non-trivial AFM driven by strong correlation enables for the spatial coexistence
of metallic regions with long-range ordered magnetization.
We suggest that this effect should also be realized at the interface
between a trivial AFM insulator and a TI. The proximity of the two phases could allow for an atomically thin AFM metallic state.
This heterostructure could be of relevance for the emerging research
field of AFM spintronics \cite{MacDonald3098,Jungwirth2018NP}.

\section*{Acknowledgements}
We thank J.C.~Budich for useful discussions and suggestions as well as the critical reading
of the manuscript. We thank M.Vojta for interesting discussions.
A.A. thanks M.Fabrizio for helpful discussions.  
A.A. and M.C. acknowledge support from the Seventh Framework
Programme FP7, under Grant No. 280555 ``GO FAST'', and the H2020
Framework Programme, under ERC Advanced Grant No. 692670
``FIRSTORM''.
A.V. acknowledges financial support from the Austrian Science Fund
(FWF) through the Erwin Schr\"odinger fellowship J3890-N36.
A.A., A.V. and M.C. also acknowledge financial support from MIUR PRIN 2015
(Prot. 2015C5SEJJ001) and SISSA/CNR project "Superconductivity,
Ferroelectricity and Magnetism in bad metals" (Prot. 232/2015).
G.S. and B.T. acknowledge financial support by the DFG (SPP 1666 on
``Topological Insulators'' and SFB 1170 ``ToCoTronics'').

\bibliography{references}
\end{document}